\documentclass{ws-ijmpa}
\usepackage[super,compress]{cite}
\usepackage{graphicx}

\begin{document}

\markboth{Prashant Shukla, Sundaresh Sankrith}{Energy and angular distributions of atmospheric muons at Earth}

\title{Energy and angular distributions of atmospheric muons at the Earth} 

\author{Prashant Shukla}

\address{Nuclear Physics Division, Bhabha Atomic Research Centre, Mumbai 400085, India.\\
Homi Bhabha National Institute, Anushakti Nagar, Mumbai 400094, India.\\
pshukla@barc.gov.in}

\author{Sundaresh Sankrith}

\address{Department of Physics, School of Advanced Sciences, Vellore Institute of Technology, Vellore 632014, Tamil Nadu, India.\\
sankrithsundaresh@gmail.com}

\maketitle

\begin{history}
\received{Day Month Year}
\revised{Day Month Year}
\end{history}

\date{\today}

\begin{abstract}
   A fair knowledge of the atmospheric muon distributions at Earth is a prerequisite
for the simulations of cosmic ray setups and rare event search detectors.
  A modified power law is proposed for atmospheric muon energy distribution which
gives a good description of the cosmic muon data in low as well as high energy regime.
  Using this distribution, analytical forms for zenith angle ($\theta$)
distribution are obtained. Assuming a flat Earth, it leads to the  $\cos^{n-1}\theta$
form where it is shown that the parameter $n$ is nothing but the power of the energy 
distribution. Exact analytical function is obtained for inclined trajectory of muon. 
A new closed form for zenith angle distribution is obtained without assuming
a flat Earth and which gives an improved description of the data at all
angles even above $70^o$. 
 These distributions are tested with the available atmospheric muon data of energy
and angular distributions. The parameters of these distributions can be used to 
characterize the cosmic muon data as a function of energy, angle and altitude.


\end{abstract}

\ccode{PACS numbers:95.85.Rj,98.70.Sa}


\section{Introduction}

  The primary cosmic rays consisting of protons, alpha particles and heavier nuclei  
continuously bombard the Earth 
at the top of the Earth atmosphere \cite{INTRO}. 
   Most of the cosmic rays originate in Galactic sources such as neutron stars, pulsars, 
supernovae and active galactic nuclei. The relative abundance of nuclei with 
charge number $Z >1$ in cosmic rays is similar to that in the interstellar medium which 
indicates that cosmic rays are the normal interstellar matter in astrophysical process. 
The majority of cosmic rays from a few GeV to 100 TeV are accelerated in supernova blast. 
 The magnetic field of the Sun tends to exclude lower energy particles (E$\approx 1$ GeV). 
During periods of low solar activity, more cosmic rays manage to reach the Earth. Earth's 
magnetic field also tends to exclude lower energy particles. The particles have greater 
difficulty penetrating the Earth's magnetic field near equator than the poles.
 Thus, the intensity of cosmic rays depends both on the location and the time.
  
   Upon entering the atmosphere, the primary cosmic radiations interact with the 
air molecules (mainly oxygen and nitrogen nuclei) mostly at 10-15 km above the 
sea level~\cite{manual}. 
  All particles suffer energy losses through hadronic and/or electromagnetic 
processes. The most abundant particles emerging from the energetic hadronic collisions 
are pions. Particles such as kaons, hyperons, charmed particles and nucleon-antinucleon pairs
are also produced.
 However, most of these particles interact further with the air or decay to other lighter particles.
 The charged pions and kaons decay to muons and neutrinos as shown below  

\begin{center}
$\pi^{-} \rightarrow \mu^{-} + \bar{\nu_{\mu}}$ \hspace{0.2in}  $\pi^{+}\rightarrow \mu^{+} + \nu_{\mu} $

$K^{-}\rightarrow \mu^{-} + \bar{\nu_{\mu}}$ \hspace{0.2in} $K^{+}\rightarrow \mu^{+} + \nu_{\mu}$ 
\end{center}
  For a given particle propagating in the atmosphere, the probabilities of decay and
interaction  become a function of energy, altitude and zenith angle. 
  The atmospheric length increases from vertical to inclined direction and this means more 
energy loss of the particles which results in a smaller integrated muon flux at the surface of 
the Earth. The pion decay probability is also larger in the inclined direction which
will enhance the muon flux in the intermediate energy.
  As the energy increases, the fraction of muons from kaon decays also increases because 
the longer-lived pions become more likely to interact before decaying than the 
shorter-lived kaons. 
  Since muons are produced with $\nu_\mu$($\overline\nu_\mu$), the measurement
of muons near the maximum intensity curve for the parent pions can be used
to calibrate the atmospheric neutrinos.

 Numerical calculations are needed to account accurately for decay and energy 
loss processes along with the knowledge of primary cosmic spectrum and the
energy dependencies of their interaction cross-sections.
 Simple parametrizations have always been useful to characterize the measured muon
momentum spectra \cite{Allkofer:1971qr,Ayre:1973df}.
  There are transport models treating the propagation of muons which can be used
to numerically obtain the muon spectra at different depths
of atmosphere \cite{Maeda:1973nz,Stephens:1979id,muon_high}.
 Detailed theoretical calculations have been performed to obtain both atmospheric muons 
and neutrino spectra at sea level \cite{Lipari:1993hd,Gaisser2002}. 
 In addition, Monte Carlo codes such as CORSIKA \cite{corsika1988} are available
which use different models to calculate the interactions of cosmic
particles in the Earth's atmosphere.
   The goal of the present paper is to obtain simple analytical expressions to provide
quick checks on how the muon distributions at Earth behave and which can be readily used
as input in the simulations of detectors for cosmic rays
as well as for rare event searches.

  In this work, a modified power law is assumed for the cosmic (atmospheric) 
muon energy distribution at Earth.
  Using this function and geometrical considerations, 
we obtain analytical form for zenith angle distribution. For a flat Earth,
it leads to $\cos^{n-1}\theta$ form where $n$ is the same as the power of energy
distribution. Further, a new analytical form for zenith angle distribution is obtained 
without assuming a flat Earth.
 The parameters of these distributions are obtained with the help of 
measured energy and angular distributions of atmospheric muons.

\section{Energy and angular distribution of atmospheric muons}

  Muons are produced at about 10-15 km height in the atmosphere and lose about 
2 GeV of energy before reaching the ground. Their energy and angular distribution 
at ground reflect a convolution of production spectrum, energy loss in the atmosphere
and the decay. 
  The energy spectrum of muons is almost flat below 1 GeV and then steepens 
to reflect the primary energy spectrum in the 10-100 GeV range.
It steepens further above 100 GeV since the pions above this energy 
would interact in the atmosphere before decaying to muons. Above 1 TeV, the 
energy spectrum of the muons is one power steeper than the primary spectrum \cite{Archard2004}.

 The energy distribution of primary cosmic rays follow power law $E^{-n}$. 
The pion and the muon distributions also follow the same power law which is 
modified in the low energy region.  The vertical flux as a function of energy 
can be described by 
\begin{equation}
I(E,\theta=0)= I_{0}\,N\, (E_{0}+ E)^{-n},
\label{flux}
\end{equation}
where $I_{0}$ is the vertical ($\theta=0$) muon flux integrated over energy,
which gives the normalization $N=(n-1)(E_0+E_c)^{(n-1)}$, where $E_c$ is the cut-off value of the data. Here, we have added a 
parameter $E_{0}$ which accounts for energy loss due to both 
the hadronic as well as electromagnetic interactions with air molecules.
  We can introduce one more parameter $\epsilon$ which modifies the power in the 
high energy part and that should account for the finite life time of pions and kaons
\begin{equation}
I(E)=  I_{0}\,N\, ( E_{0} + E)^{-n} \left(1 + \frac{E}{\epsilon}\right)^{-1}.
\label{flux1}
\end{equation}
Both the Eqs.~\ref{flux} and \ref{flux1} assume that the energy loss ($E_0$) is 
independent of particle energy, an assumption which is good for minimum ionizing particles. 
 At low energies, the energy loss varies as $1/E$ thus 
a more appropriate distribution would come with an additional parameter $E_1$ as
\begin{equation}
I(E)=  I_{0}\,N\, \left( E_{0} +E_1/E + E\right)^{-n} \left(1 + \frac{E}{\epsilon}\right)^{-1}.
\label{flux2}
\end{equation}
Here, the normalization constant $N$ can be obtained numerically. However, we will 
not use the Eq.~\ref{flux2} for the analysis presented in this paper.



\begin{figure}
\begin{center}
\includegraphics[width=0.60\linewidth]{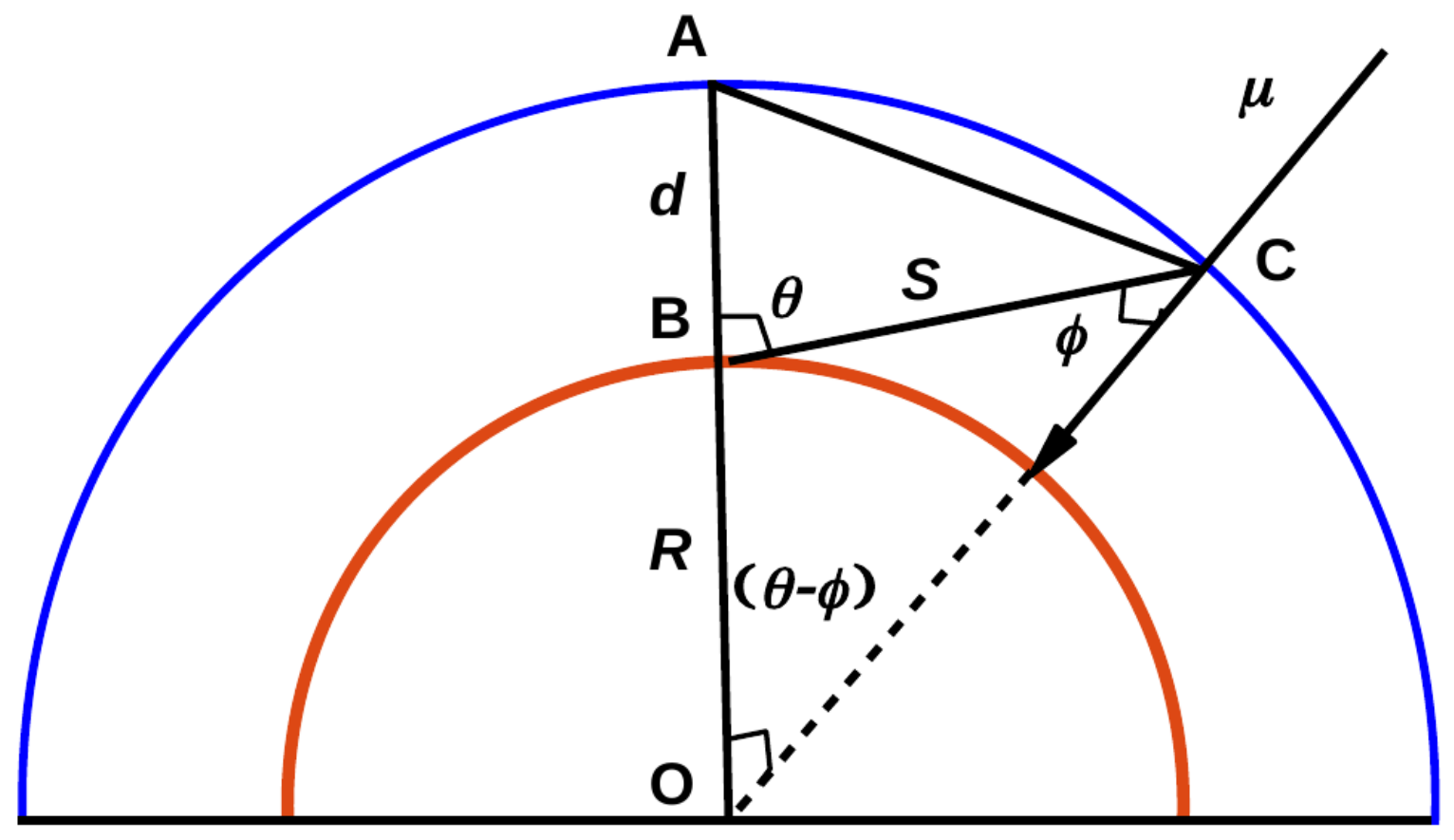}
\caption{Geometrical relation between the vertical pathlength $d$ and the pathlength 
inclined at a zenith angle $\theta$.}
\label{rbd}
\end{center}
\end{figure}

\begin{figure}
\begin{center}
\includegraphics[width=0.60\linewidth]{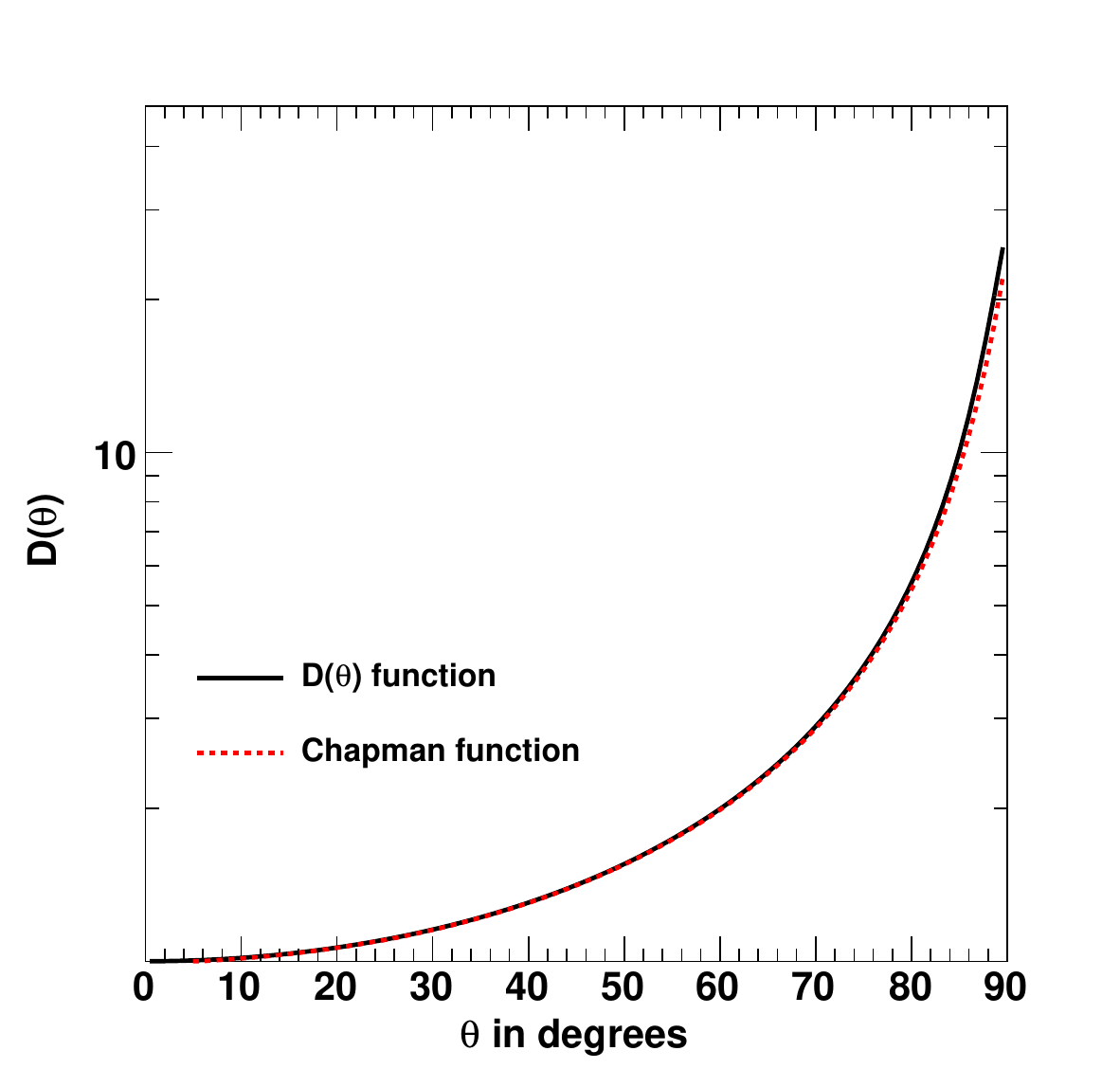}
\caption{Comparison of the analytical expression for the ratio of
  inclined to vertical pathlengths of muon in Eq.~\ref{Dtheta} with
  the Chapman function~\cite{manual}.}
\label{Chapman}
\end{center}
\end{figure}

The muon flux measured on the Earth's surface has a weak dependence on the azimuthal angle 
but depends strongly on Zenith angle which is defined as the angle made by the incident 
ray with the vertical direction at that point.
  One can obtain inclined distance $S$ in terms of vertical distance $d$, 
the zenith angle $\theta$ and the Earth's radius $R$ using a 
simple geometrical picture given in Fig.\ref{rbd}.
  Using sine law for $\bigtriangleup OBC$ 

\begin{equation}
\frac{\sin\phi}{R} = \frac{\sin(\theta-\phi)}{S} = \frac{\sin(180-\theta)}{R+d}.
\label{ratio1}
\end{equation}
The relation between $\phi$ and $\theta$ is 

\begin{equation}
  \sin\phi=\frac{R}{R+d} \, \sin\theta
\label{ratio2}
\end{equation}
and the pathlength $S$ in the inclined direction is

\begin{equation}
  S=\frac{\sin(\theta-\phi)}{\sin(\theta)} (R+d).
\label{ratio4}
\end{equation}
Using Eq.~\ref{ratio2} and \ref{ratio4}, the ratio of pathlengths of
a muon from inclined direction to that of a muon from the vertical direction
is obtained as
\begin{equation}
D(\theta) =  \frac{S}{d} = \sqrt{\left(\frac{R^{2}}{d^{2}}\cos^2\theta + 2 \frac{R}{d} + 1\right)} 
          - \frac{R}{d} \cos\theta.
\label{Dtheta}
\end{equation}
Equation~\ref{Dtheta} provides a closed expression for column density for an inclined
trajectory of muons in curved Earth's atmosphere and is a replacement of Chapman's function
given in the first chapter of the book by Grieder~\cite{manual}. 
Accuracy of the various approximations to the Chapman function is discussed by Swider 
and Gardner \cite{Swider1967}.
Figure~\ref{Chapman} shows the comparison of the analytical expression for the ratio of
inclined to vertical pathlengths of muon in Eq.~\ref{Dtheta} with the Chapman function.
 
 The ratio of the integrated muon flux at $\theta$ with that at $0^{o}$ 
can be obtained from Eq.~\ref{flux} as


\begin{equation*}
\frac{\Phi(\theta)}{\Phi(\theta=0)}  
    = \frac{\int_{0}^{\infty}  \left(E_0 + E_{\theta}+E\right)^{-n} \,dE} 
                   {\int_{0}^{\infty}  \, (E_0 + E )^{-n} \, dE}
\end{equation*}

\begin{equation}
 = \left( \frac{E_{0}+E_{\theta}}{E_{0}} \right)^{-(n-1)}
\end{equation}
where $E_{0}+E_{\theta}$ is the energy loss of the muon in the inclined direction. The integrated 
flux from Eq.~\ref{flux1} can also be considered the same as above since the higher energy 
term depending on $\epsilon$ will make a little difference to the integrated flux. 
  The ratio of the energy loss from inclined to the vertical direction is given by 
the ratio of the pathlengths (same as the ratio of thicknesses) $D(\theta)$ in the respective 
directions and thus, the zenith angle distribution of energy integrated flux in terms 
of $I_0 = \Phi(\theta=0)$ is obtained as 
\begin{equation}
 \Phi(\theta) = I_{0} \, D(\theta)^{-(n-1)}.
 \label{newfit}
\end{equation}
 The overall distribution function as a function of both the energy and the zenith angle then 
can be written as 
\begin{equation}
  I(E,\theta)= I_{0} \, N \, (E_0 + E )^{-n} \, \left( {1 + \frac{E}{\epsilon}}\right) ^{-1} \, \, 
                    D(\theta)^{-(n-1)}. 
\label{allDist}
\end{equation}
  Here, the function $D(\theta)$ is given by Eq~\ref{Dtheta}. 
If the Earth is assumed to be flat then $D(\theta) = 1/\cos\theta$ which on putting
in Eq.~\ref{newfit} leads to  
\begin{equation}
\Phi(\theta)= I_0 \, \cos^{n-1}\theta.
\label{oldfit}
\end{equation}
 With $n\simeq 3$, this gives the usual $\cos^2\theta$ distribution which is widely 
used to describe Zenith angle distribution. This expression gives good
description of the data at lower Zenith angle but not at higher angles because
it assumes a flat Earth.

  Gaisser had given the formula \cite{INTRO} for muon energy distribution assuming
flat Earth and which is valid for high energy ($E_{\mu} > 100/\rm cos\theta$ GeV)
which is written as 
\begin{equation}
\frac{dN_{\mu}}{dE d\Omega}  \approx {1400E_{\mu}^{-2.7}} / (\rm{m^{2} s GeV sr}) 
      \left(\frac{1}{1+ \frac{1.1E\,\rm cos\theta}{\epsilon_{\pi}}} 
        + \frac{0.054}{1+\frac{1.1 E\, \rm cos\theta} {\epsilon_{K}}}\right),
\label{Gaisser}
\end{equation}
 where the two terms in the bracket give the contributions of pions and kaons
in terms of two parameters $\epsilon_{\pi} \approx 115$ GeV and 
$\epsilon_{K} \approx 850 $ GeV.
  In the distribution function given by Eq.~\ref{allDist} we use only one parameter 
$\epsilon$ which is obtained by fitting the experimental data.

\section{Analysis of the measured data}
\begin{figure}
\begin{center}
\includegraphics[width=0.65\textwidth]{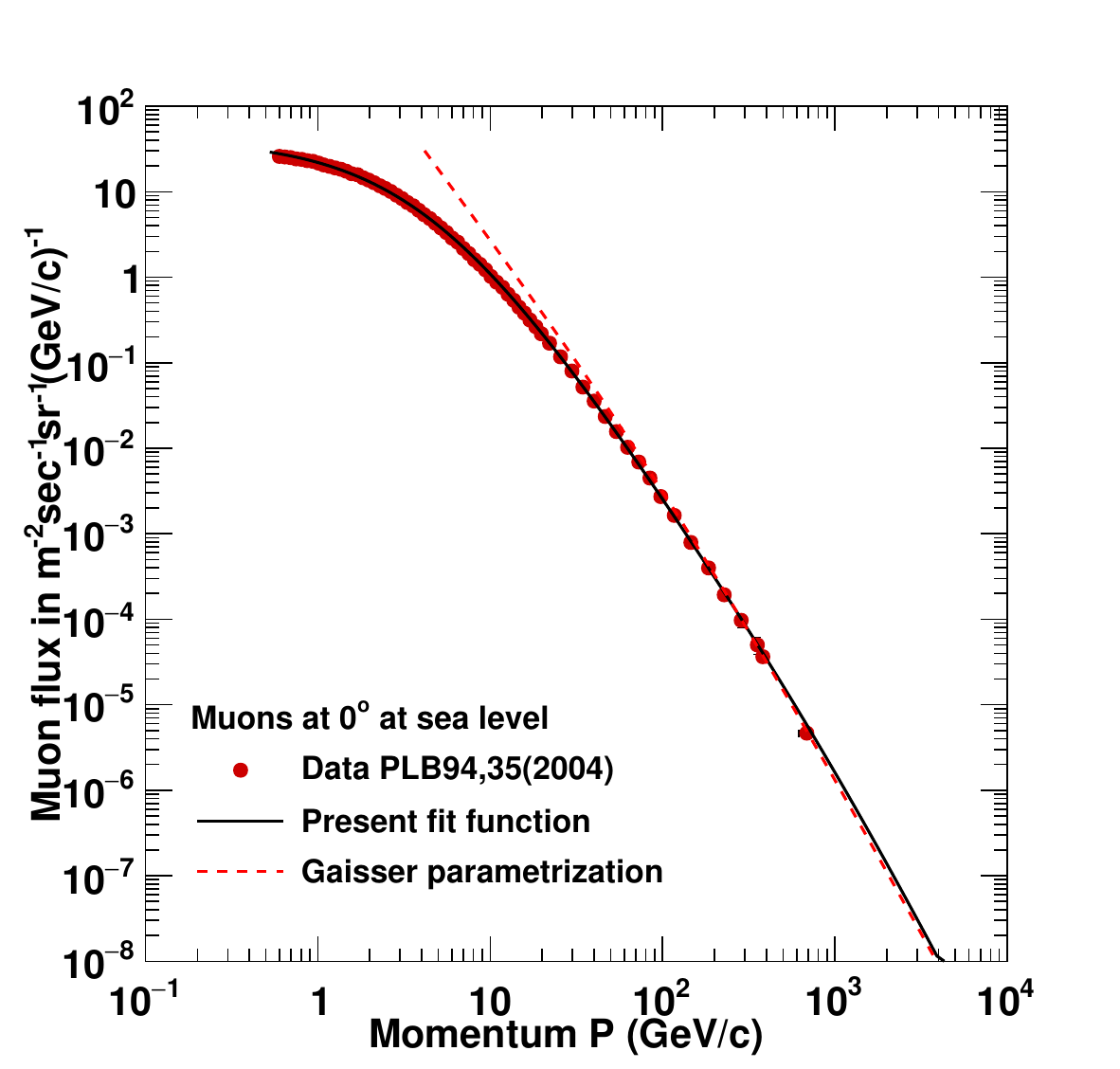}
\caption{Muon momentum distribution at $0^{o}$ zenith angle at sea level~\cite{Haino}. 
 The lines show the fits with Eq.~\ref{flux1} and Eq.~\ref{Gaisser}.}
\label{muonZero}
\end{center}
\end{figure}

\begin{figure}
\begin{center}
\includegraphics[width=0.65\textwidth]{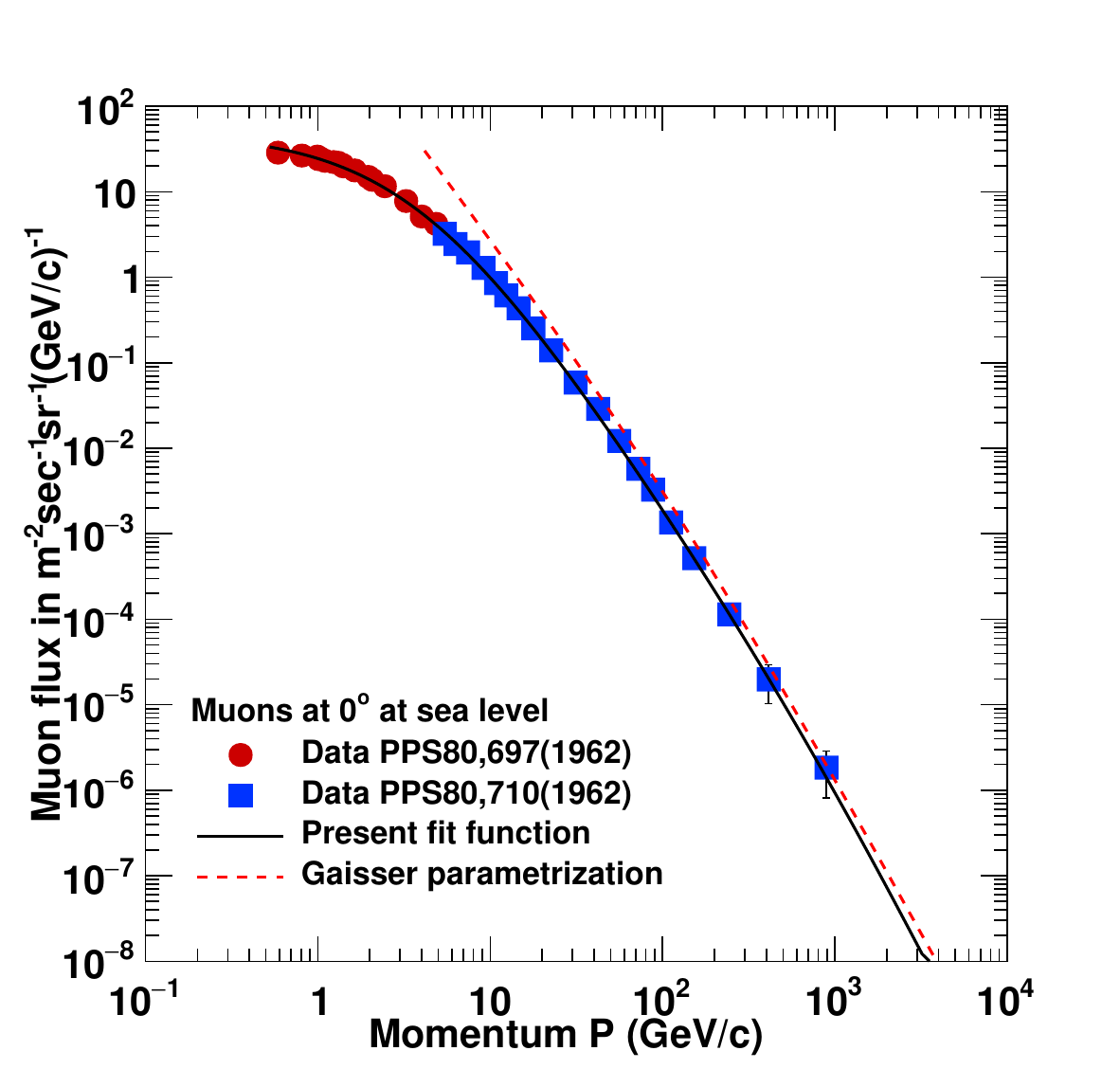}
\caption{Muon momentum distribution at $0^{o}$ zenith angle at sea
  level~\cite{0370-1328-80-3-314,0370-1328-80-3-315}. 
 The lines show the fits with Eq.~\ref{flux1} and Eq.~\ref{Gaisser}.}
\label{muonZero2}
\end{center}
\end{figure}

\begin{figure}
\begin{center}
\includegraphics[width=0.65\textwidth]{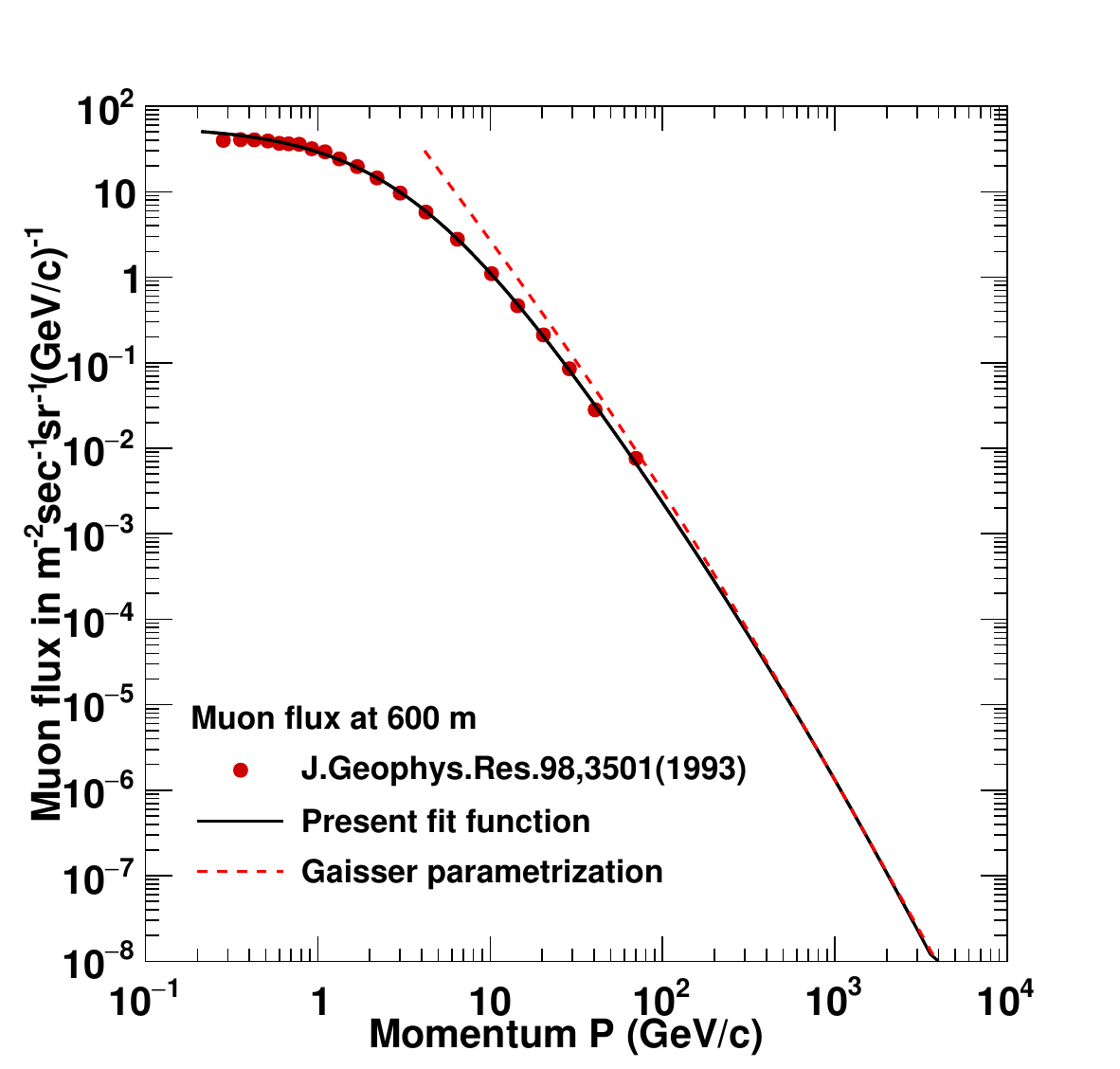}
\caption{Muon momentum distribution at $0^{o}$ zenith angle at 600 m 
 altitude~\cite{h600}. The lines show the fits with Eq.~\ref{flux1} and 
 Eq.~\ref{Gaisser}.}
\label{muon600}
\end{center}
\end{figure}

  We choose four datasets of muon energy distribution namely at sea level,
at high altitude and at an inclined angle and fit them with 
the function given in Eq.~\ref{flux1} to obtain the parameters.
 We also analyse primary cosmic ray (protons and helium) distributions using the 
same Eq.~\ref{flux1}.
  Figures~\ref{muonZero} and~\ref{muonZero2} show the momentum distributions of 
atmospheric muons at $0^{o}$ zenith angle at sea level measured at two different locations
Tsukuba~\cite{Haino} and Durham~\cite{0370-1328-80-3-314,0370-1328-80-3-315}, respectively.
Figure~\ref{muon600} shows the same at 600 m altitude~\cite{h600}. The lines show the fits with 
Eq.~\ref{flux1} and Eq.~\ref{Gaisser}. The Gaisser function gives good description of the data 
at high momentum
The present function gives excellent description of both the low as well as the
high momentum part of the muon distribution and thus the parameter $I_0$ gives a
reliable estimate of the integrated flux at $0^{o}$ zenith angle.

\begin{figure}
\begin{center}
\includegraphics[width=0.65\textwidth]{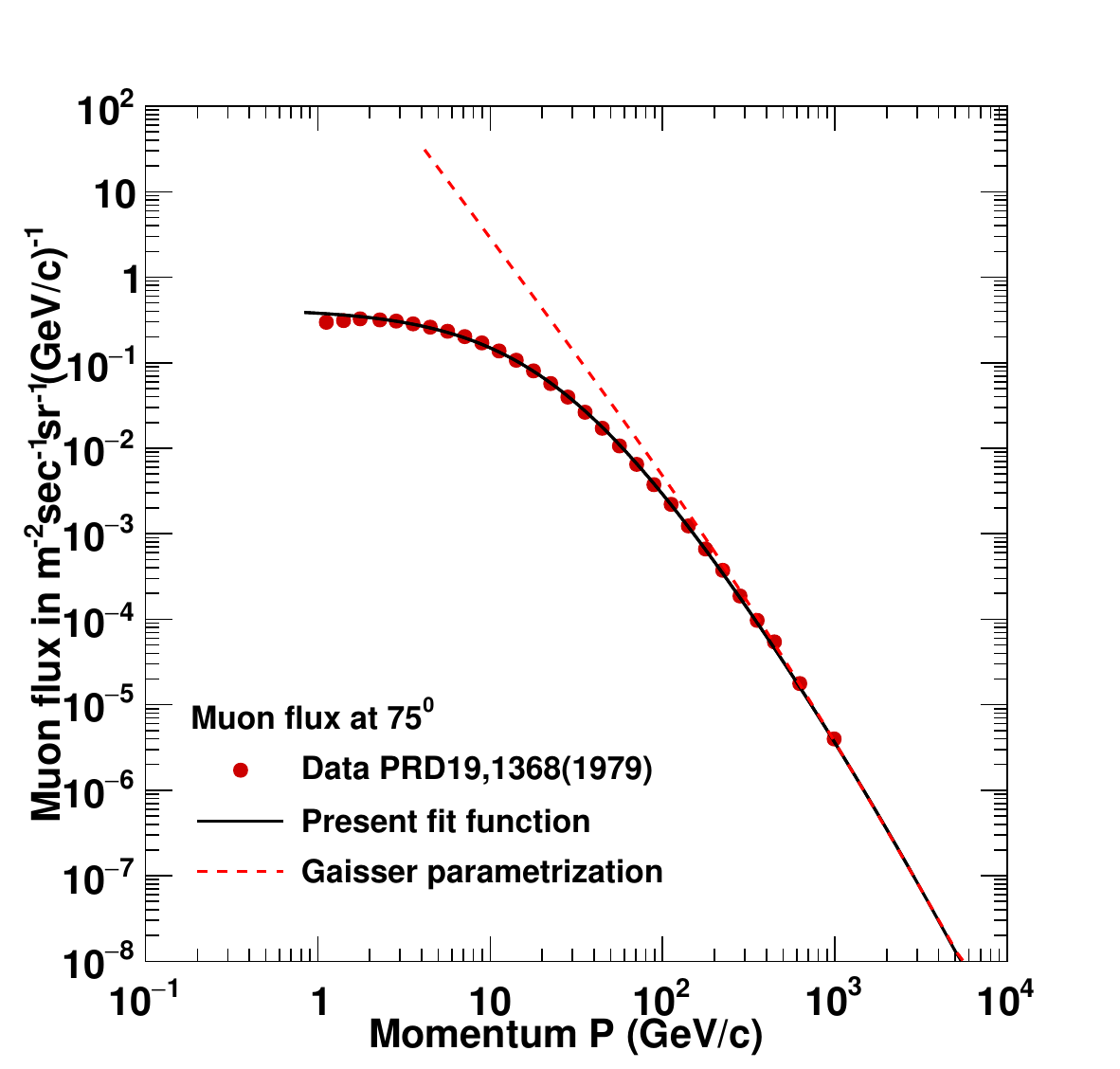}
\caption{Muon momentum distribution at sea level measured at Zenith angle $75^{o}$~\cite{ang75} 
fitted with Eq.~\ref{flux1} and Eq.~\ref{Gaisser}.}
\label{muon75}
\end{center}
\end{figure}

Figure~\ref{muon75} shows the muon momentum distribution at sea level but measured at 
Zenith angle $75^{o}$~\cite{ang75} fitted with Eq.~\ref{flux1} and Eq.~\ref{Gaisser}.
The function in Eq.~\ref{flux1} describes the data well though there is
an expected mismatch at the lowest momentum.

\begin{figure}
\begin{center}
\includegraphics[width=0.65\textwidth]{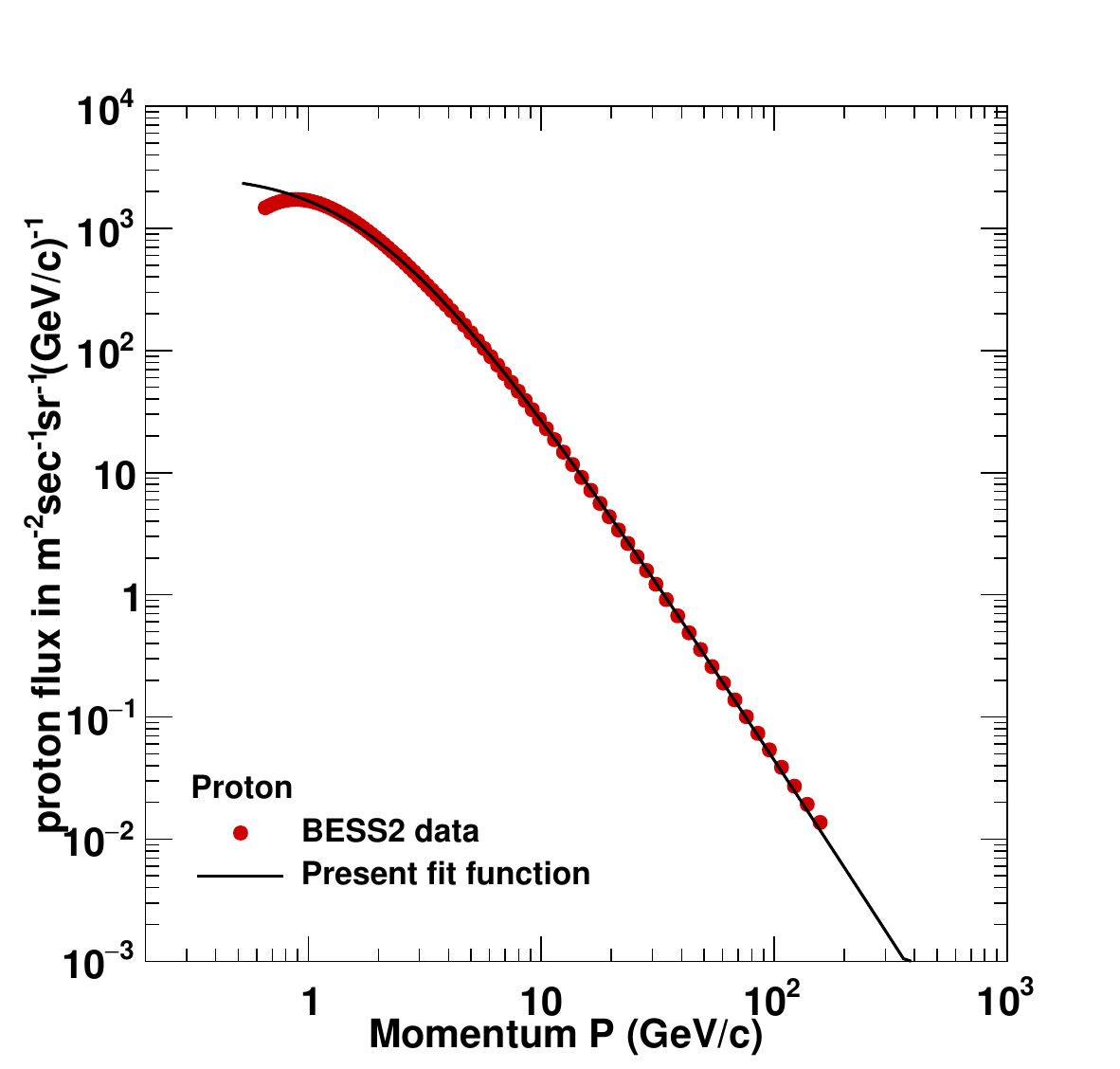}
\caption{Proton flux ~\cite{protonPrim} as a function of momentum at the 
 top of the atmosphere fitted with Eq.~\ref{flux1}.}
\label{Figproton}
\end{center}
\end{figure}

\begin{figure}
\begin{center}
\includegraphics[width=0.65\textwidth]{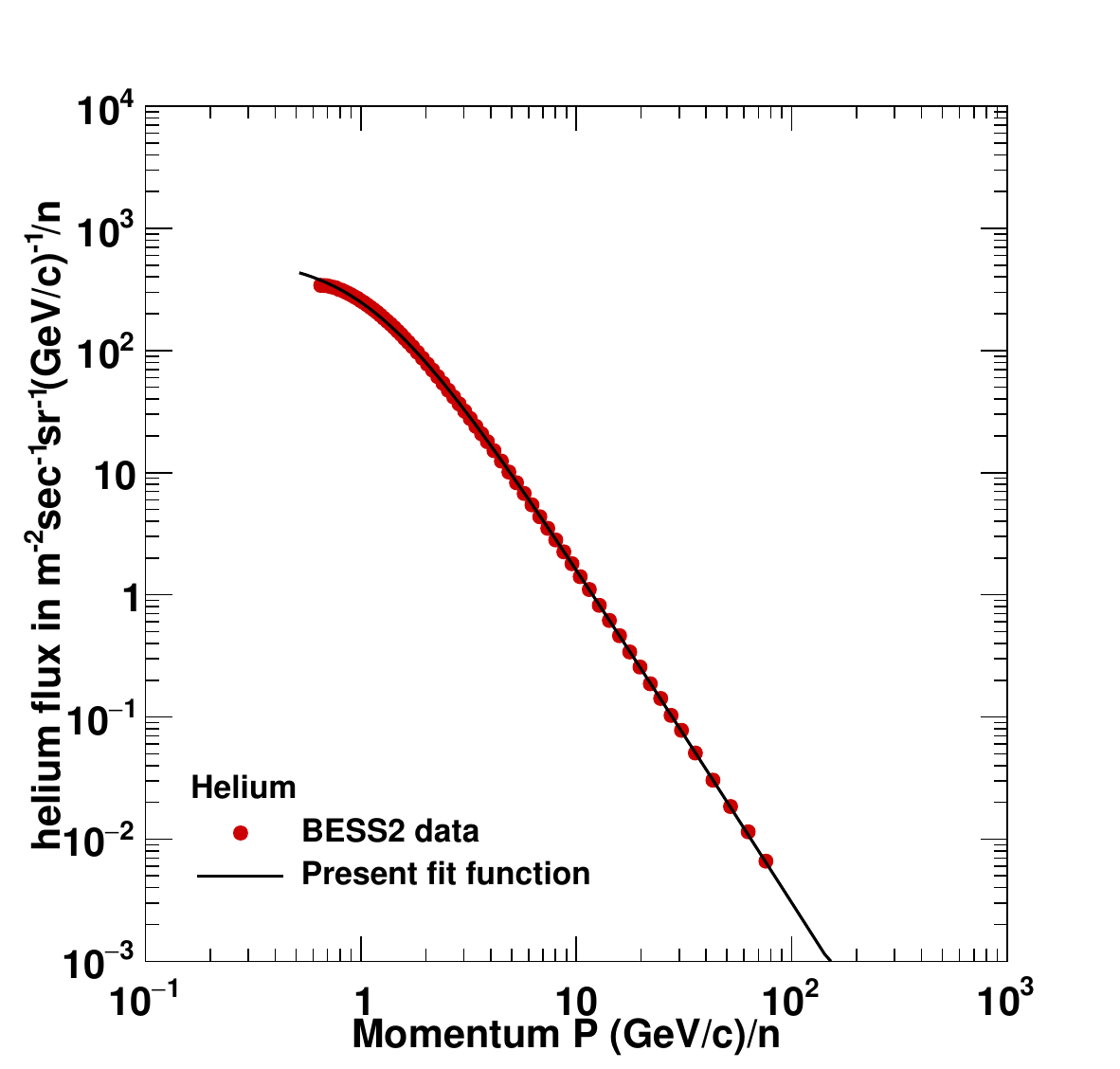}
\caption{Helium flux~\cite{protonPrim} as a function of momentum at the 
 top of the atmosphere fitted with Eq.~\ref{flux1}.}
\label{FigHe}
\end{center}
\end{figure}

Figure~\ref{Figproton} shows Proton flux and and the Fig.~\ref{FigHe} shows the 
Helium flux~\cite{protonPrim} as a function of momentum at the  top of the atmosphere 
fitted with Eq.~\ref{flux1}. The aim here is to get the shape and the power $n$ for the
primary cosmic spectra. There are recent measurements of primary cosmic particles
from PAMELA detector \cite{Adriani:2013xva} in the high momentum range.

 The values of the fit parameters corresponding to all the data analysed 
are listed in Table~\ref{Epara1}.
   The value of the power $n$ of the energy distribution is around 3 for 
muons at sea level, at 600m altitude and at an inclined angle. For protons
$n=2.93$ and for Helium it is 2.75 which means that the muon spectra become
slightly steeper than the primary rays due to the interaction processes in 
the atmosphere.
 The integrated flux $I_0$ at $\theta=0$ is $72.5\pm 0.2$ m$^{-2}$s$^{-1}$sr$^{-1}$
at sea level which  increases to $98.8\pm 0.5$ m$^{-2}$s$^{-1}$sr$^{-1}$ at 600 m.
The value of the parameter $I_0$ obtained at a zentih angle of $75^{o}$ at sea level
is $65.2\pm 2$ m$^{-2}$s$^{-1}$sr$^{-1}$.  
 The value of parameter $E_{0}$ for muons is 4.29 GeV at ground and becomes
smaller at 600 m above the ground. 
  The value of $E_0$ is very high for muons at $75^{o}$ due to longer 
pathlength in the atmosphere. For proton and helium its value is small but 
finite, showing the interactions before they are detected. 
  The parameter $\epsilon$ is $854\pm 105$ for vertical flux. For the other datasets 
it is fixed so as to have an agreement with the Gaisser distribution since there
is no data in the high energy region to constrain this parameter.


\begin{figure}
\begin{center}
\includegraphics[width=0.65\textwidth]{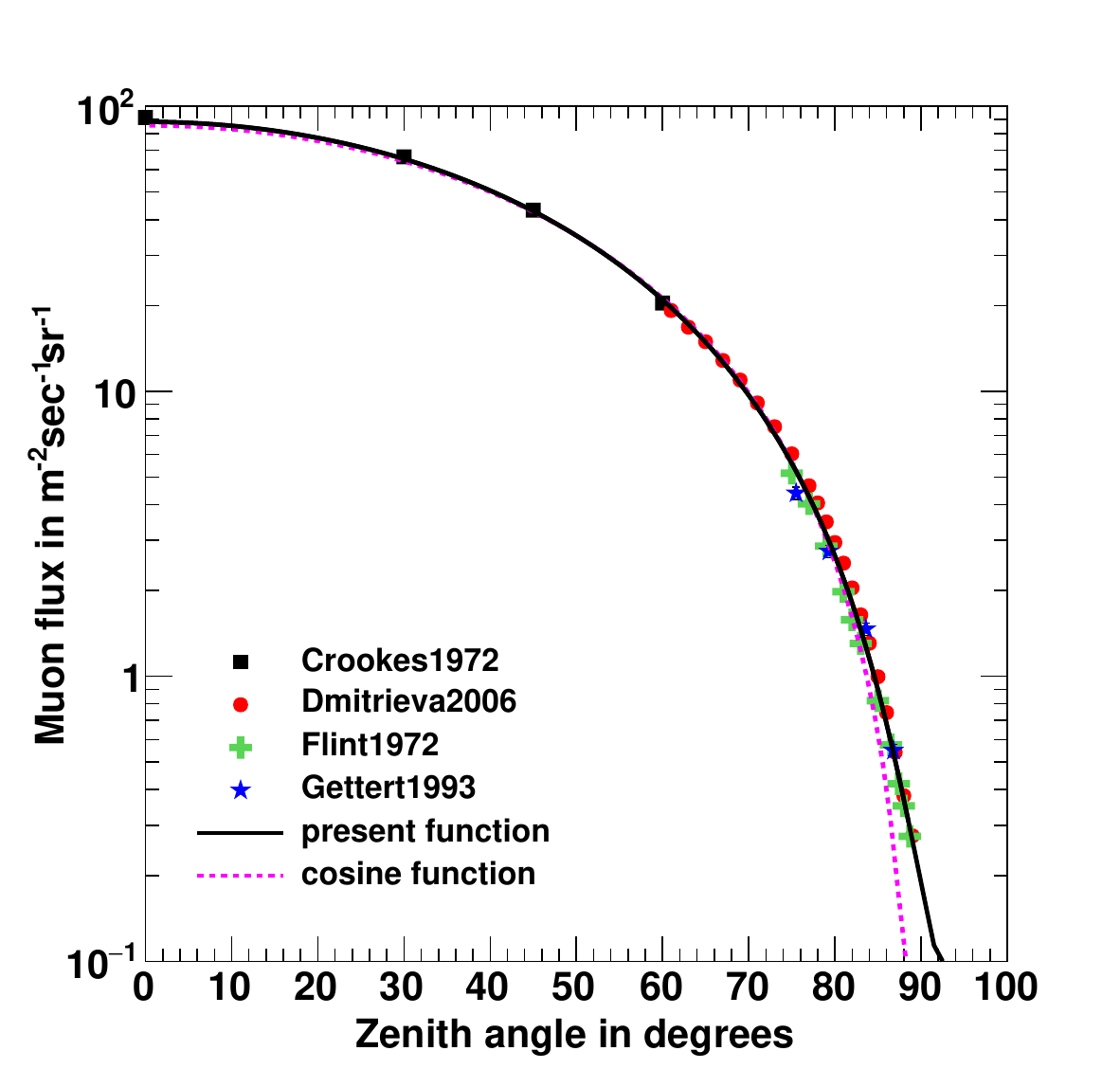}
\caption{Muon flux as a function of zenith angle~\cite{zenith} at sea level fitted 
with Eq.~\ref{newfit} and Eq.~\ref{oldfit}.}
\label{FigZenith}
\end{center}
\end{figure}

\begin{table}[ph]
\tbl{Parameters of Eq.~\ref{flux1} obtained by fitting the measured muon distributions.}
{\begin{tabular} {@{}ccccccc@{}} \toprule
\hline
             & $I_{0}$  (m$^{-2}$    & $n$   & $E_{0}$    & $1/\epsilon$   &$\chi^{2}/ndf$&  Data Reference  \\
             & s$^{-1}$ sr$^{-1})$   &       &   (GeV)   &  (GeV)$^{-1}$   &   &   \\          
\hline
\hline
 $\mu$ at $0^{\circ}$  & 70.7     & 3.01      & 4.29      & 1/854       & 128/63 &  Tsukuba, Japan  \\ 
   sea level         & $\pm$0.2  & $\pm$0.01  & $\pm$0.04  &          & &  ($36.2^o$ N, $140.1^{o}$ W) \\ 
    (E$>$0.5 GeV)        &            &            &            &          & &                      \\  \hline
 $\mu$ at $0^{\circ}$  & 72.5      & 3.06      & 3.87      & 1/854     & 60.8/23  &  Durham, UK  \\ 
   sea level        & $\pm$0.2  & $\pm$0.03  & $\pm$0.07  &  (fixed) &       &  ($54.76^o$ N, $1.57^{o}$ W) 
\\
(E$>$0.5 GeV)        &            &            &            &          & &                      \\ 
\hline
  $\mu$ at $0^{\circ}$ & 98.8      & 3.00     & 3.6        & 1/854       & 60.4/18  & Prince Albert, Canada      \\
  at 600 m           & $\pm$0.5     & $\pm$0.08  & $\pm$0.1   &  (fixed)  &  &  ($53.2^{o}$ N, $105.75^{o}$ W)    \\
  (E$>$0.2 GeV)        &            &            &            &          & &                      \\  
\hline
 $\mu$ at $75^{\circ}$ & 65.2     & 3.00     &  23.78     & 1/2000     &  48.3/25   & Hamburg, Germany   \\
  sea level       & $\pm$1.5     & $\pm$0.02  & $\pm$0.30  & (fixed)     &  & ($53.56^{o}$ N, $10^{o}$ E)   \\ 
  (E$>$1.0 GeV)        &            &            &            &          & &                      \\ 
\hline
  Proton             & 5000       &  2.93       & 1.42     & 0.0      & 96.3/86    & Antarctica    \\ 
 (E$>$0.5 GeV)       & $\pm$52    & $\pm$0.01  & 0.04     &         &     &   \\
\hline
  Helium          & 863      &  2.75       & 0.28       &  0.0     & 20.7/67 & Antarctica     \\  
 (E$>$0.5 GeV)   & $\pm$15   & $\pm$0.02   & $\pm$0.03  &          &   &    \\ 
\hline
\end{tabular}
\label{Epara1}}
\end{table}






  The value of the parameter $R/d$ is fixed at 174.0 which is obtained by fitting the
parameters of Eq.~\ref{newfit} with the zenith angle distributions 
measured by various experiments shown in Fig.~\ref{FigZenith}.
  The data are taken from the collection of Ref.~\cite{zenith} with the original 
references \cite{Crookes1972,Dmitrieva2006,Flint1972,Gettert1993}. The different 
datasets have different muon energy thresholds and we take the normalized data from the 
review \cite{zenith}. Ideally, we should have a dataset from a single experiment 
covering large range of zenith angles. 

 Table~\ref{Epara2} lists the parameters obtained from the measured Zenith Angle 
distribution.
  The fit with the function $\cos^{n-1}\theta$  has been restricted 
below $80^0$. With $n\sim3$ this gives us the popular $\cos^{2}\theta$ distribution.
The present distribution Eq.~\ref{newfit} gives excellent description 
of the data at all angles. 
  The parameter $I_0$ obtained from the present distribution and the 
$\cos^{n-1}\theta$ distribution match with each other.
 Another observation is the value of the power $n$ obtained from the energy 
distribution is very close to the value obtained from fitting the zenith angle 
distribution. This is the most important result of this study.

\begin{table}[h]
\tbl{Parameters obtained from the measured Zenith Angle distribution.}
{\begin{tabular} {@{}ccccc@{}} \toprule
\hline
       Fit function   &  $I_{0}$    & $n$   &  $R/d$      & $\chi^{2}/ndf$    \\ 
\hline                
  $\Phi(\theta) = I_0 \, D(\theta)^{-(n-1)}$ & 88.0$\pm$2.4   &  3.09$\pm$0.03 &  174$\pm$12  & 111/37  \\ 
\hline 
  $\Phi(\theta) = I_0 \, \cos^{(n-1)}\theta$ & 85.6$\pm$2.4 &  3.01$\pm$0.03  &  -     & 52/17   \\ 
\hline
\end{tabular}
\label{Epara2}}
\end{table}

\clearpage

\section{Conclusions}

  In this work, analytical functions are proposed for muon energy and
angle distributions. A modified power law gives a good description of the cosmic 
muon momentum distribution in low as well as high energy region.
  Using the modified power law form of energy distribution, analytical forms for 
zenith angle distribution are obtained. Assuming a flat Earth, it leads to 
the  $\cos^{n-1}\theta$ form where it is shown that the parameter $n$ is nothing but the 
power of the energy distribution. With $n\sim3$ it leads to the famous $\cos^{2}\theta$
distribution. Exact analytical function is obtained for inclined trajectory of muon.
  A new analytical form for zenith angle distribution is obtained without
assuming a flat Earth which gives an excellent description of the data at 
all zenith angles. 
  These functions explain the shape of the spectra and are useful to get the 
integrated flux. Their parameters are useful to characterize the data 
as a function of energy, angle and altitude.

\clearpage


\end{document}